\documentclass[%
 preprint,
 superscriptaddress,
 amsmath,amssymb,showkeys,
 aps,
]{revtex4-1}

\usepackage{graphicx}
\usepackage{dcolumn}
\usepackage{bm}
\usepackage{hyperref}
\usepackage[mathlines]{lineno}
\usepackage{array}
\usepackage{booktabs}

\begin{document}
\title{Understanding the mesoscopic scaling patterns within cities}

\author{Lei Dong}
\affiliation{
 Institute of Remote Sensing and Geographical Information Systems, School of Earth and Space Sciences, Peking University, Beijing 100871, China
}
\affiliation{Senseable City Lab, Department of Urban Studies and Planning, Massachusetts Institute of Technology, Cambridge, MA 02139, USA}

\author{Zhou Huang}
\affiliation{
 Institute of Remote Sensing and Geographical Information Systems, School of Earth and Space Sciences, Peking University, Beijing 100871, China
}

\author{Jiang Zhang}
\affiliation{
 School of System Science, Beijing Normal University, Beijing 100875, China
}

\author{Yu Liu}
\email{liuyu@urban.pku.edu.cn}
\affiliation{
 Institute of Remote Sensing and Geographical Information Systems, School of Earth and Space Sciences, Peking University, Beijing 100871, China
}

\begin{abstract}
Understanding quantitative relationships between urban elements is crucial for a wide range of applications. The observation at the macroscopic level demonstrates that the aggregated urban quantities (e.g., gross domestic product) scale systematically with population sizes across cities, also known as urban scaling laws. However, at the mesoscopic level, we lack an understanding of whether the simple scaling relationship holds within cities, which is a fundamental question regarding the spatial origin of scaling in urban systems. Here, by analyzing four extensive datasets covering millions of mobile phone users and urban facilities, we investigate the scaling phenomena within cities. We find that the mesoscopic infrastructure volume and socioeconomic activity scale sub- and super-linearly with the active population, respectively. For a same scaling phenomenon, however, the exponents vary in cities of similar population sizes. To explain these empirical observations, we propose a conceptual framework by considering the heterogeneous distributions of population and facilities, and the spatial interactions between them. Analytical and numerical results suggest that, despite the large number of complexities that influence urban activities, the simple interaction rules can effectively explain the observed regularity and heterogeneity in scaling behaviors within cities.
\end{abstract}

\maketitle

\section*{Introduction}
In spite of the complexity and variety of cities, it turns out that various macroscopic properties related to urban activities $Y$, such as gross domestic product and infrastructure, scale with the population size $P$ in a surprisingly simple power-law manner: $Y \sim P^{\beta}$, where $\beta$ is a scaling exponent (or an elasticity, in economic terms) that characterizes the non-linear properties of urban systems \cite{bettencourt2007growth}. In the past decades, the macroscopic urban scaling phenomena have drawn great scientific interest in physics \cite{popovic2012geometric,yakubo2014superlinear,barthelemy2016structure}, economics \cite{youn2016scaling,gomez2017explaining}, transportation \cite{samaniego2008cities,louf2014congestion}, environment \cite{ribeiro2019effects}, urban studies \cite{batty2011defining,li2019singularity}, and interdisciplinary fields \cite{batty2008size,bettencourt2010unified,bettencourt2013origins}. And data in many urban systems have demonstrated that these power-law relationships remain remarkably stable in different countries \cite{bettencourt2007growth,li2017simple} and historical periods \cite{hamilton2007nonlinear,ortman2015settlement}.

At the mesoscopic level, however, whether the relationships between urban characteristics obey some universal patterns remains poorly understood. Here, the notion of the mesoscopic level means a spatial scale around a few kilometers within cities, which is the most commonly used spatial unit for urban research and urban planning \cite{hall2014cities}. Moreover, a striking variation in population and socioeconomic density emerges at this spatial scale \cite{ahlfeldt2019economic,timothy2001intra,duranton2004micro}. Nevertheless, current urban scaling frameworks `ignore' those heterogeneous distributions as they usually model a city as a whole and study the macroscopic scaling phenomena across cities \cite{bettencourt2013origins,pan2013urban,um2009scaling,gomez2017explaining,ribeiro2019effects,keuschnigg2019urban} or the temporal dynamics of individual cities \cite{keuschnigg2019scaling,depersin2018global,ribeiro2020relation}. (Ref. \cite{bettencourt2020interpretation} compares the cross-sectional and temporal scaling analyses at the macroscopic level.) Several key questions at the mesoscopic level remain unanswered: do sub-units within a single city follow the power-law scaling as observed for systems of cities? What is the mechanism behind the potential scaling patterns within cities? Answering these questions is critical to reach a better understanding of urban systems. 

\begin{figure*}[!htbp]
    \centering
    \includegraphics[width=1.\linewidth]{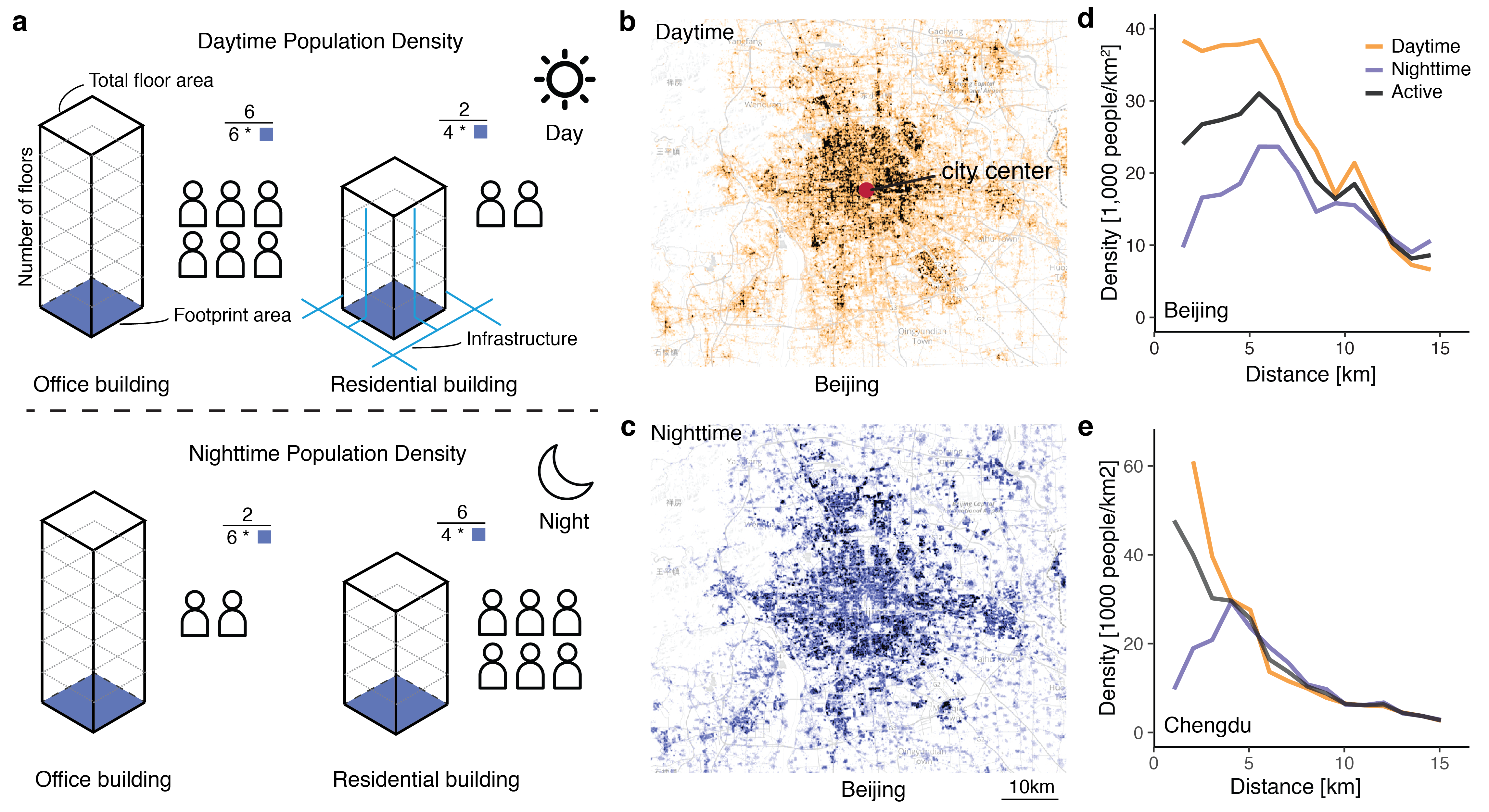}
    \caption{\textbf{Spatio-temporal dynamics of population.} \textbf{a}, Illustration of the temporal dynamics of population. The office building is of high population density in the daytime and of low density in the nighttime. The residential building, however, is the opposite. We assume that for each building the infrastructure volume (blue lines) is proportional to total building areas, which equals footprint area times the number of floors. \textbf{b, c}, The spatial distributions of daytime (\textbf{b}) and nighttime (\textbf{c}) populations for Beijing. These maps were created with the Datamaps tool (https://github.com/ericfischer/datamaps). The base map is OpenStreetMap, which is licenced under the Creative Commons Attribution-ShareAlike 2.0 licence (CC BY-SA). \textbf{d, e}, The daytime, nighttime, and active population density gradients from the city center to the periphery for Beijing (\textbf{d}) and Chengdu (\textbf{e}). As expected, the daytime population density is much higher than the nighttime density around the urban center in all studies cities. Statistically, exponential decay function has a higher $R^2$ for both daytime and nighttime populations in Beijing. Power decay function achieves higher for daytime density, and exponential function has the better performance for nighttime density in Chengdu.}
    \label{fig:1}
\end{figure*}

Our limited understanding of intra-urban scaling phenomena stems from the lack of granular data documenting the spatial distributions of urban elements such as population, infrastructure, and socioeconomic activity. Meanwhile, increasing urban dynamics presents further challenges to the data and measurement issue \cite{strumsky2019different}. For instance, population -- the key urban element -- is quite dynamic within cities, making it `inaccurate' when measuring population distribution by static data like census data. As the census population only reflects a snapshot of the nighttime distribution of residents, the daytime density of the urban center is highly underestimated (Fig.~\ref{fig:1} and Supplementary Fig. 1). Recently, researchers have taken crucial steps in mapping the dynamic population \cite{deville2014dynamic} or considering three-dimensional building morphologies \cite{batty2008scaling,schlapfer2015urban,liu2018vertical} in the within-city analysis. Nevertheless, quantitative relations between urban elements are still far from clear. 

Here, benefiting from the revolution of big data, we analyze the quantitative relationships between population, infrastructure, and socioeconomic activity at the mesoscopic level of ten Chinese cities: Beijing, Chengdu, Hangzhou, Jinan, Nanjing, Shanghai, Shenzhen, Suzhou, Xi'an, and Zhengzhou (Supplementary Table 1). These cities locate in different geographic regions of China, which helps test the robustness of our findings. To derive the quantitative relationship between urban elements (Table~\ref{tab:1}), we use four extensive micro-datasets, including a granular mobile phone dataset covering 107 million people, a building dataset containing the three-dimensional information of $\sim$ 2 million buildings, a firm dataset recording $\sim$ 13 million firms, and a point of interests (POIs) dataset with approximately 1 million commercial facilities, see Methods for detailed data descriptions. The mobile phone data allow us to construct an `active population' measure to capture the population dynamics (detailed below); and the building data provide the venue to quantify the three-dimensional development of infrastructure (Fig.~\ref{fig:1}a). Based on these datasets, we have three empirical observations. First, we find a robust sub-linear relationship between active population and infrastructure volume and a robust super-linear relationship for socioeconomic activity within cities. Second, the average intra-urban scaling exponents are consistent with the empirical and theoretical results across cities. Third, the exponents of different cities, however, are also notably different. 

To explain these observations, we propose a conceptual framework that unifies the heterogeneous population distribution and the spatial interactions between people-infrastructure and people-people. We decompose spatial interactions into two effects. The \textit{local effect} captures the interaction between local population density and infrastructure networks. The \textit{global effect} captures the city-wide interactions between population via the gravity equation, and the spatial distribution of the active population is regarded as a two-dimensional gravity field. Analytical and numerical results suggest, despite the large number of complexities that influence urban activities, the simple spatial interaction rules can effectively predict sub- and super-linear scaling behaviors within cities. The interaction intensity, a city-specific parameter we introduced in each rule, can explain the difference in scaling exponents. These findings offer a mechanistic understanding of scaling phenomena within cities \cite{arcaute2015constructing}, and echo the fractal and self-similar nature of cities \cite{batty1994fractal}.

\section*{Results}
\subsection*{Active population} 
To incorporate the temporal dynamics and derive a better measure of the population distribution, we employ the concept of the \textit{active population} (AP), which is a more appropriate proxy than simply residential or employment population for estimating socioeconomic activity \cite{li2017simple}. The AP reflects a mixture of the daytime and nighttime populations within a given region by combing them together with the active time as a weight $\lambda$:
 
\begin{equation}
    AP = \lambda P_{day} + (1-\lambda) P_{night}.
\label{eq1}
\end{equation}

Here, the daytime and nighttime populations are estimated by a large-scale mobile phone dataset for the year 2015 (Fig.~\ref{fig:1} and Methods). For the ten cities studied, we have a total of 107 million mobile phone users, accounting for more than 75\% of the total population of these cities (see Supplementary Table 1 for details).

The total AP in one city is the same as the total daytime or nighttime population if there is no intercity commuting. Stated simply, we further assume that the ratio of daytime and nighttime is approximately 1:1, i.e., 12 hours for daytime and 12 hours for nighttime in one day. Therefore, we have $\lambda = 1/2$ in Eq.~(\ref{eq1}). In other words, here we use the average of the daytime and nighttime population as a measure of the AP. One benefit of this setting is that for cities without mobile phone data, AP could be calculated by the employment (daytime) and residential (nighttime) populations, which are available in many cities' official statistics. In Supplementary Fig. 2, we further show the results by adjusting $\lambda$ within reasonable limits ($\lambda = 1/3$ and $2/3$), and all conclusions are robust.

We present the daytime, nighttime, and active population density gradients from the downtown to the urban fringe of Beijing and Chengdu in Fig.~\ref{fig:1}de. Previous studies have found that population density decays from the city center with an exponential, power-law-like, or some more complex forms \cite{martori2016real}. We find similar patterns in the granular population data. The population density curves, however, vary significantly between day and night, especially around the urban core areas as shown in Fig.~\ref{fig:1}de.

\begin{table}[!htbp]
    \centering
    \begin{tabular}{*3c}
    \hline
        Characteristics & Proxy variables & Dataset\\
    \hline
        Population & Active population & Mobile phone dataset\\
        Infrastructure volume & Total floor area of buildings & Building dataset \\
        Socioeconomic activity (interaction) &  Number of firms  & Firm dataset \\
         & POIs of commercial facilities & POIs dataset\\
    \hline
    \end{tabular}
    \caption{Variables used to estimate scaling characteristics.}
    \begin{flushleft}
    Note: We assume that the proxy variable is proportional to the corresponding scaling characteristics.
    \end{flushleft}
    \label{tab:1}
\end{table}

\begin{figure*}[!htbp]
    \centering
    \includegraphics[width=1.\linewidth]{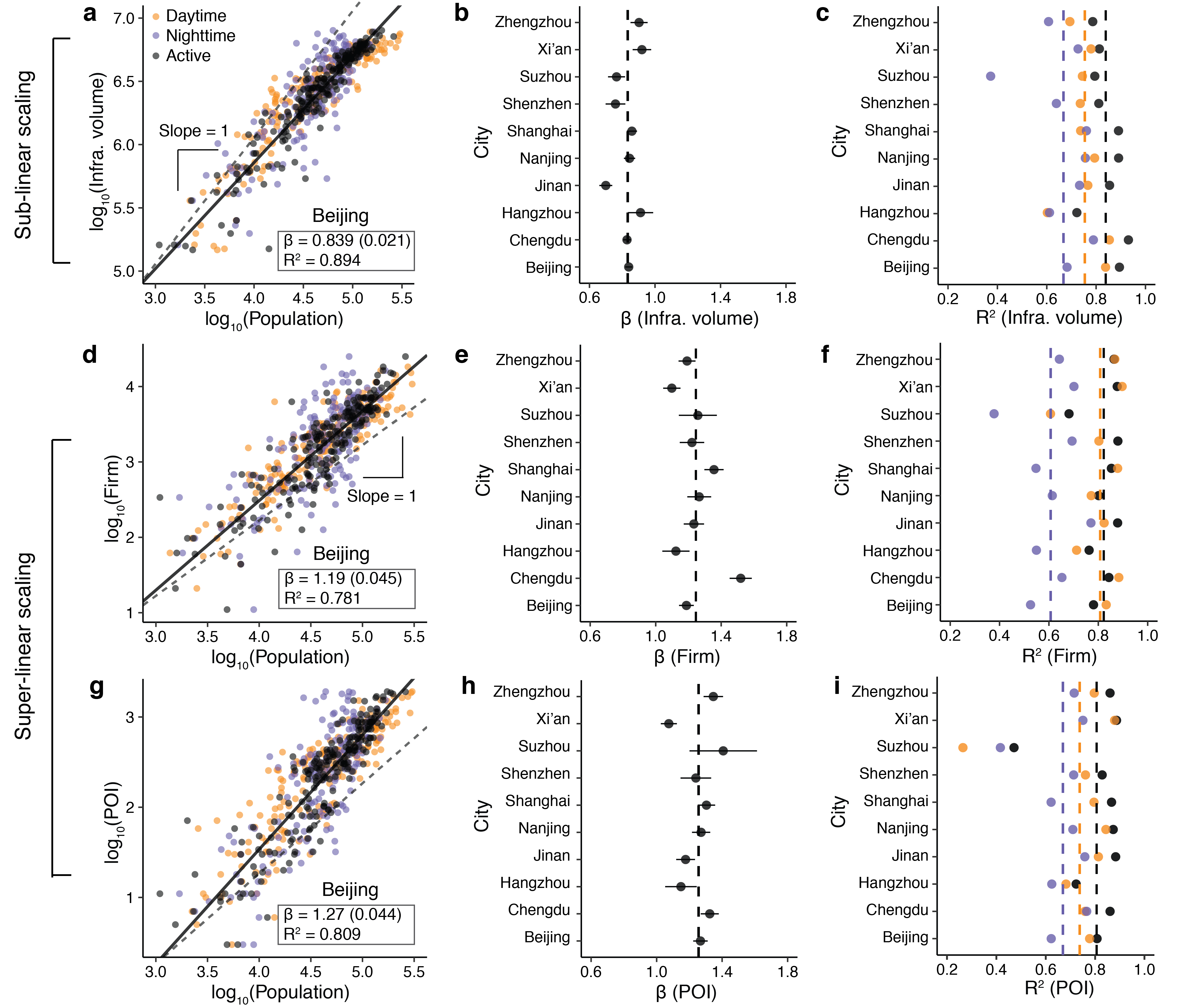}
    \caption{\textbf{Intra-urban scaling of infrastructure and socioeconomic interactions.} \textbf{a-c}, The sub-linear scaling between population and infrastructure volume. \textbf{d-i}, The super-linear scaling between population and the number of firms (d-f) and the number of POIs (g-i). \textbf{a, d, g}, The scatter plots and fitting results of Beijing for the infrastructure volume (a), the number of firms (d), and the number of POIs (g). \textbf{b, e, h}, The scaling exponents ($\pm$ one standard error) of ten studied cities. \textbf{c, f, i}, $R^2$ of daytime, nighttime, and active populations. The mean values of $\beta$ and $R^2$ are labeled with dashed lines.}
    \label{fig:2}
\end{figure*}

\subsection*{The empirical findings}
Given the detailed spatial distributions of urban elements, a proper spatial unit is then required to perform the statistical analysis. To make the results of different cities comparable, here we use a $2km \times 2km$ grid as our analysis units (Methods). To address the potential modifiable areal unit problem (MAUP), meaning the statistical results are influenced by the scale of the aggregation unit \cite {openshow1979million,briant2010dots,louf2014scaling}, we also perform a robustness check by varying the grid size, and all results are stable (Supplementary Table 2 and Supplementary Fig. 3). 

We aggregate the daytime/nighttime/active populations, buildings, firms, and POIs into the corresponding grid cell. To derive the scaling exponent, we take the simplest fitting procedure: 

\begin{equation}
    \log_{10} Y_i = \log_{10} Y_0 + \beta \log_{10} P_i + \epsilon_i,
\end{equation}

\noindent where $i$ indexes different grid cells in a city, the dependent variable $Y_i$ denotes the infrastructure volume, the number of firms, or the number of POIs, and $P_i$ is the population size. $\epsilon_i$ is the error term. 

The fitting results between population and infrastructure volume are shown in Fig.~\ref{fig:2}a-c. Here we use the total building areas (i.e., building volumes) to represent the infrastructure volume by assuming a linear relationship between them (for example, one elevator services a certain amount of building areas in office buildings or apartments; see Table~\ref{tab:1}). The reason to use buildings rather than road networks to measure the infrastructure volume is that within cities, especially in high-density areas, a large amount of infrastructure is vertically distributed. Road networks would underestimate infrastructure volumes in these areas.

Figure~\ref{fig:2}b shows that in all studied cities, the scaling exponents of infrastructure are less than 1, indicating a robust sub-linear relationship with the population size. Interestingly, the average value (over all cities, the dashed line in Fig.~\ref{fig:2}b) between active population and infrastructure volume $\langle \beta_{infra | ap} \rangle \approx 0.833$ is very close to 5/6, a theoretical value of the scaling exponent between infrastructure and population across cities \cite{bettencourt2013origins}. Moreover, Figure \ref{fig:2}c clearly shows that compared with daytime and nighttime populations, the AP achieves the highest $R^2$ in all cities ($\langle R^2_{infra | ap} \rangle \approx 0.839$), which demonstrates the effectiveness of the AP measurement. 

To investigate the super-linear scaling within cities, we collect two granular socioeconomic activity datasets: the firm registration record data and the POI data (see Methods). We use the number of firms and POIs as the proxy variables for socioeconomic activity (Table~\ref{tab:1}). Figure~\ref{fig:2}eh shows that the super-linear scaling between AP and socioeconomic activity holds well in both datasets. In all ten cities, the scaling exponents of firms and POIs are both greater than 1, and the average value is approximately 1.25, which is very close to the empirical results across cities and the theoretical values of 7/6 \cite{bettencourt2013origins} or 4/3 \cite{zhang2015scaling} derived from different models. Similar to the infrastructure results, the $R^2$ calculated by the AP is the highest in most cities (Fig.~\ref{fig:2}fi). We notice that for the firm dataset, the daytime population also performs well in terms of the $R^2$. This is not difficult to understand, as most firm-related activities occur during the day and are closely related to the daytime population (employment) distribution. 

Despite the robust sub/super-linear relationships, we can also observe differences in the scaling exponents (Fig.~\ref{fig:2}). Specifically, for the same scaling phenomenon, the scaling exponents of cities with similar population sizes can be statistically different. For instance, the population of Shenzhen and Xi'an is similar (between 12 and 13 million). However, the exponents of infrastructure, number of firms and POIs in the two cities are significantly different (Fig.~\ref{fig:2}). A similar pattern is found in the data of Beijing and Shanghai (population is between 22 and 24 million), firms scale more superlinearly in Shanghai compared with Beijing (Fig.~\ref{fig:2}e). These findings suggest that population size is not the only determining parameter that influences the scaling phenomena within cities.

\subsection*{The conceptual framework} 
\begin{figure*}[!htbp]
    \centering
    \includegraphics[width=1.\linewidth]{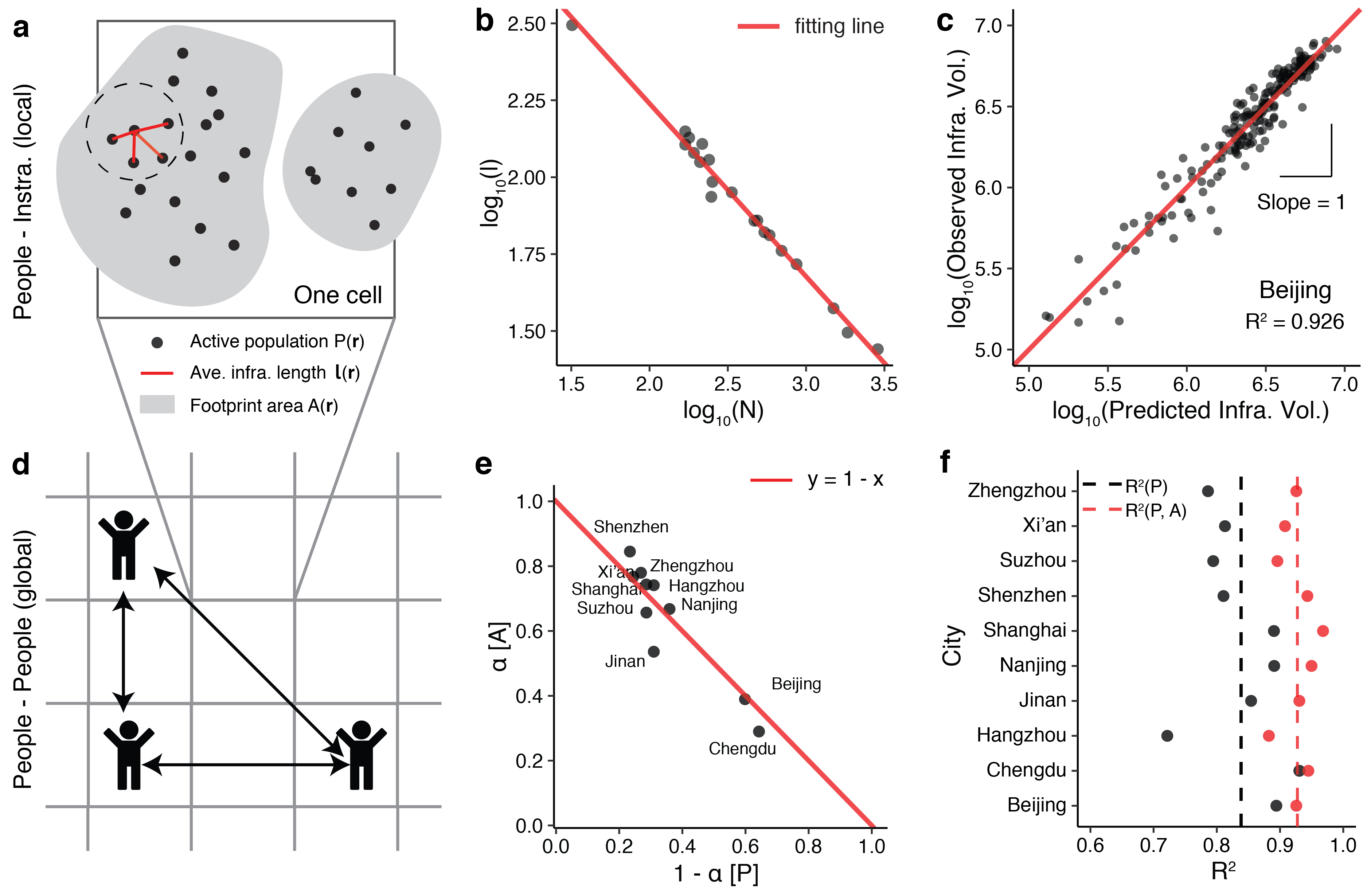}
    \caption{\textbf{Interactions and sub-linear scaling.} \textbf{a,} Illustration of the localized connection between AP and infrastructure. We assume that an AP connects to its $n$ nearest neighbors by the infrastructure network, and $n$ is a constant number. For simplicity, we draw a two-dimensional schematic. In fact, the population distribution and infrastructure connections are three-dimensional. \textbf{b}, $\ell \sim \rho^{-\alpha}$ of road network data ($\alpha$ = 0.562, $R^2$ = 0.985). \textbf{c}, Scatter plot between the observed and predicted infrastructure volumes obtained from Eq. (\ref{rho}). \textbf{d,} Illustration of the global interaction between people and people. \textbf{e,} Scatter plots of $\langle 1-\alpha \rangle$ and $\langle \alpha \rangle$, which are the exponents of $P$ and $A$ in the regression $\log_{10} V_i = C_i + (1-\alpha) \log_{10} P_i + \alpha \log_{10} A_i$, respectively. The red line is the prediction of the Cobb-Douglas function with constant returns to scale. \textbf{f,} $R^2$s of the ten studied cities. The average $R^2$ obtained from Eq.~(\ref{ipa}) is 0.927 (red dots), and we also put the results of Fig.~\ref{fig:2}c here (black dots) for comparison.}
   \label{fig:3}
\end{figure*}

To explain these empirical observations simultaneously, we propose a conceptual framework. The main ideas are that the two key elements that constitute a city, its physical infrastructure and socioeconomic activity, can be modeled by the local and global spatial interactions with its citizens, respectively. The sub-linear scaling is derived by local interactions between population and infrastructure (Fig.~\ref{fig:3}a), because infrastructure networks develop in a decentralized way in order to connect people \cite{bettencourt2013origins}. The super-linear scaling is assumed to be the results of global interactions between population (Fig.~\ref{fig:3}d). All of our analyses below consider the heterogeneous population distribution, and this goes beyond previous theoretical frameworks, which usually assume a uniform population distribution \cite{bettencourt2013origins,pan2013urban}.

Let $\rho_i$ denote the population density of cell $i$, and $\rho_i = P_i / A_i$, where $P_i$ is the active population and $A_i$ is the building footprint area within cell $i$ (gray areas in Fig.~\ref{fig:3}a). $A_i$ is derived by summing the footprint area of all buildings located in cell $i$. Since infrastructure services population in a localized way, we assume that the typical length of infrastructure (e.g., roads, pipes, and cables) $\ell$ depends on $\rho$ in the following form

\begin{equation}
    \ell\sim \rho^{-\alpha},
\label{rho}
\end{equation}
\noindent
where $\alpha$ ($0 < \alpha < 1$) is a city-specific parameter controlling the local interaction intensity. This equation can be verified with empirical data of road networks (Fig.~\ref{fig:3}b), simulated results (Supplementary Fig. 4), and some spatial network models \cite{barthelemy2008modeling}. Specifically, we collect road network data from \cite{cardillo2006structural}, which includes twenty 1 square mile samples of different world cities. We then calculate the correlation between the average road length $\ell$ and the density of road intersections (a proxy for population density as the number of road intersections is proportional to the population size \cite{strano2012elementary}). Figure~\ref{fig:3}b shows that these two variables are well-fitted by Eq.~(\ref{rho}), indicating that our assumption is backed up by empirical data. The total infrastructure length $V_i$ within cell $i$ is thus given by the product of the population $P_i$ and the average infrastructure length $\ell_i$

\begin{equation}
    V_i = \ell_i P_i = (P_i / A_i)^{-\alpha} P_i \sim  P_i^{1-\alpha}A_i^{\alpha}.
\label{ipa}
\end{equation}

This equation means that the larger the $\alpha$, the smaller the impact of the population $P$ on the infrastructure. We notice that Eq.~(\ref{ipa}) is a special case of the Cobb-Douglas production function \cite{cobb1928theory, ribeiro2019effects}, which displays constant returns to scale as the sum of the exponents equals 1 ($1 - \alpha + \alpha \equiv 1$). The constant returns to scale means that doubling the population $P$ and footprint area $A$ will also double infrastructure volume $V$. We take the logarithm of Eq.~(\ref{ipa}) and perform a simple OLS regression to estimate the coefficients, $(1-\alpha)$ and $\alpha$, for each city. As shown in Fig.~\ref{fig:3}e, the exponents of $P$ and $A$ of different cities almost perfectly fall on the predicted line given by the constant returns to scale property. 

The analytical and empirical results of Eq.~(\ref{ipa}) imply that both population and footprint area can contribute to infrastructure volume, which is rarely mentioned in the scaling literature. In other words, population is \textit{not} the only determining factor that affects the infrastructure within cities (similarly, Ref. \cite{ribeiro2019effects} finds that population and built-up area jointly affect the urban carbon dioxide emissions). Take some newly developed areas in a city for example, the population size of these areas has not yet grown; therefore the infrastructure volume of these areas is much higher than the value predicted by the current population. A similar issue exists for urban slums, where the infrastructure is much lower than the estimated number based on their population size \cite{brelsford2018toward}. These intra-urban variations in land use partially explain why the data points shown in Fig.~\ref{fig:2} are much noisier than the cross-city plots. By considering both $P$ and $A$ we can obtain a better fitting result for infrastructure (Fig.~\ref{fig:3}c and Supplementary Fig. 5), and the average $R^2$ increases from 0.839 to 0.927 (Fig.~\ref{fig:3}f). 

Although $P$, $A$, and $V$ are coupled together as shown in Eq.~(\ref{ipa}), we can still obtain a simple scaling exponent between $P$ and $V$ by assuming a power-law relationship between $P$ and $A$. In Supplementary Fig. 6, we empirically show that $A \sim P^{\eta}$ ($\langle \eta \rangle \approx 0.734 < 1$). Thus, we obtain

\begin{equation}
    V_i \sim P_i^{1 - \alpha (1 - \eta)}.
\label{eq.sub}
\end{equation}

The exponent $\beta_{sub} = 1 - \alpha (1 - \eta)$ is less than 1, indicating a sub-linear scaling. The tunable parameters $\alpha$ and $\eta$ capture the heterogeneity in different cities.

Unlike sub-linear scaling, we argue that the super-linear scaling within cities is the result of global (i.e., city-wide) interactions between people (Fig.~\ref{fig:3}d). To model the global interactions, we employ the gravity model, which is widely used to mimic the interaction flows (e.g., people, goods) between different regions \cite{popovic2012geometric,rybski2013distance,yakubo2014superlinear}. This practice also links urban scaling to human mobility, as the gravity model is one of the essential mobility models. 

Let $q_{ij}$ denote the interaction between cell $i$ and $j$, according to the gravity model, we have
\begin{equation}
q_{ij} = \frac{k P_i P_j}{d_{ij}^{\gamma}},
\label{eq:gravity1}
\end{equation}
\noindent
where $d_{ij}$ is the Euclidean distance between the centroid of cell $i$ and $j$, $\gamma$ is a parameter controlling the geographical constrain for the interaction, and $k$ is the constant. This equation includes two effects: i) the active population $P_i$ captures the preferential attachment meaning a popular location will attract more people; ii) $d_{ij}^{\gamma}$ captures the spatial dependence. Here, $\gamma = 1$ is particularly noteworthy because of $\gamma = 1$ exactly corresponding to the gravity field in a two-dimensional space \cite{mazzoli2019field}, and the model becomes a `parameter-free' model under this setting. Experimentally, the value of $\gamma$ ranges in the interval [1, 1.5] \cite{deville2016scaling,liu2012understanding,ribeiro2017model,disdier2008puzzling}. 

$Q_i$, the total interactions of location $i$, can be derived by summing Eq.~(\ref{eq:gravity1}): 

\begin{equation}
Q_{i} = k P_i \sum_{j \neq i} \frac{P_j}{d_{ij}^{\gamma}}.
\end{equation}

Due to the complicated spatial correlation between $P_j$ and $d_{ij}$, there is no general analytical solution for $Q_i$; here we present the numerical estimations based on the population distribution of the studied cities. Figure~\ref{fig:4}a shows interactions $Q_i$ as a function of the active population size $P_i$ for Beijing (see Supplementary Fig. 7 for the results of the remaining cities). As can be seen, all data points fall almost exactly on a straight line with a slope greater than one, indicating that the gravity function can virtually reproduce the super-linear scaling between population and interactions. More importantly, we find that $\beta_{sup}$ derived by our `parameter-free' model ($\gamma = 1$) is very close to the theoretical value 7/6 across cities \cite{bettencourt2013origins}, which provides some new insights into the long-standing debate over the gravity model coefficients in urban fields \cite{simini2012universal}. 

Figure~\ref{fig:4}b further shows that the scaling exponent $\beta_{sup}$ increases monotonically as $\gamma$ increases, and $\beta_{sup}$ ranges from 1.15 to 1.34 when $\gamma$ ranges from 1 to 2. And we find a linear relationship between $\gamma$ and $\beta_{sup}$ within this range:

\begin{equation}
    \beta_{sup} \approx a + b(\gamma - 1),
\end{equation}
\noindent
where $a = 1.153 \, (0.001)$ and $b = 0.186 \, (0.000)$ ($R^2 = 0.999$). $\beta_{sup}$ derived from the model is quite similar to our empirical findings (see Fig.~\ref{fig:2}, we assume the number of firms and POIs is proportional to the volume of interactions). Also, the tunable parameter $\gamma$ reflects the variations of global interaction in different cities and different urban phenomena.

\begin{figure*}[!htbp]
    \centering
    \includegraphics[width=.8\linewidth]{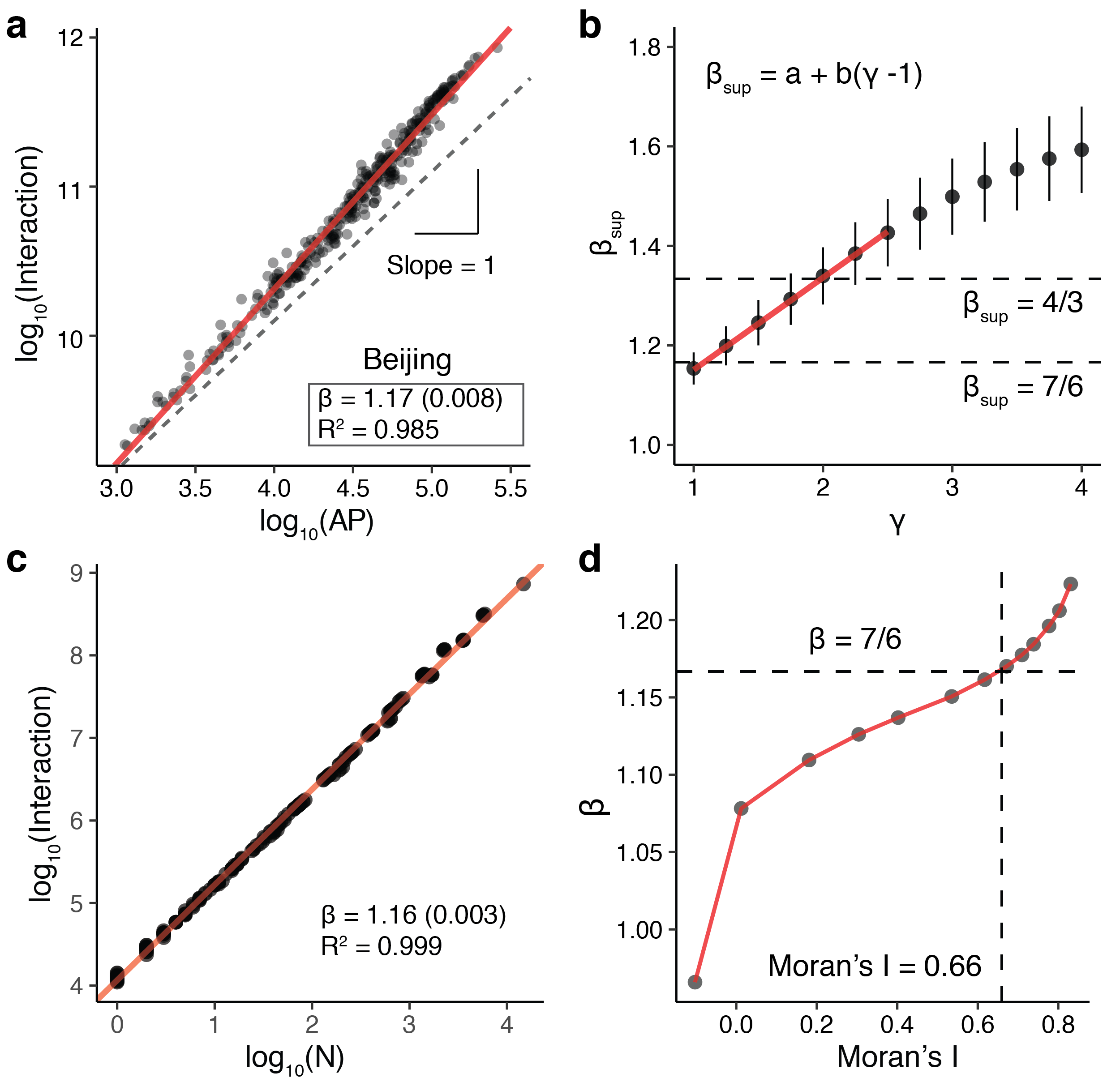}
    \caption{\textbf{Gravity model, Moran's I, and super-linear scaling.} \textbf{a}, Fitting results between AP and gravity-based interactions for Beijing ($\gamma = 1$). \textbf{b,} Urban scaling exponent $\beta_{sup}$ changes with the values of $\gamma$. The mean values of $\beta_{sup}$ (y-axis) were calculated based on the simulation results of the ten studied cities (with $\pm$ one standard deviation). Interestingly, we find a linear relationship between $\gamma$ and $\beta_{sup}$ when $\gamma$ ranges from 1 to 2 (the red line), and the cases $\gamma=1$ and $2$ effectively reproduce the theoretical estimations of $\beta = 7/6$ and $4/3$, respectively. \textbf{c,} The super-linear scaling between interaction and the number of nodes (population) with $\sigma = 1$ and $\gamma = 1$. \textbf{d,} Moran's I and the scaling exponent $\beta$. $\beta$ increases monotonically as Moran's I increases, and the theoretical value $\beta = 7/6$ corresponds to Moran's I = 0.66, the similar value to the empirical results of Moran's I.}
    \label{fig:4}
\end{figure*}

\begin{table}[!htbp]
    \centering
    \begin{tabular}{*5c}
    \hline
        Exponents & \multicolumn{2}{c}{Within cities} & \multicolumn{2}{c}{Across cities}\\
     {} & Observation & Model & Observation & Model\\
    \hline
        $\beta_{sub}$ & [0.70,0.92] & $1 - \alpha (1 - \eta)$ & [0.74,0.92] & $1-\delta$\\
        $\beta_{sup}$ & [1.07,1.41] & $a + b(\gamma-1)$ & [1.01,1.33] & $1+\delta$ \\
    \hline
    \end{tabular}
    \caption{Scaling exponents within cities and across cities}
    \begin{flushleft}
    Note: The empirical and theoretical results across cities are obtained from Ref. \cite{bettencourt2013origins}.
    \end{flushleft}
    \label{tab:2}
\end{table}

\subsection*{Spatial autocorrelation, gravity, and super-linear scaling}

We notice that the population distributions of different cities fluctuate considerably (Supplementary Fig. 1), but all cities have similar super-linear scaling exponents under the same $\gamma$ (Supplementary Fig. 7). It is supposed that there should be some unified hidden parameters behind the spatial distribution of population contributing to the universal super-linear scaling behaviors. Spatial autocorrelation is a good candidate for that parameter under the intra-urban setting, because most geographical phenomena have positive spatial autocorrelation and dependence \cite{anselin2001spatial}. To test this assumption, we perform two analyses. First, we randomly swap the values of the cell-level population; thus, everything is preserved but the spatial correlations are destroyed. Supplementary Fig. 8 shows that the scaling exponent between population and interaction becomes 1 after the shuffling, indicating that the super-linear scaling phenomenon disappears after eliminating the spatial correlations.

Second, we calculate Moran's I \cite{moran1950notes}, the most commonly used indicator for spatial autocorrelation and show the connection between Moran's I and the scaling exponent (Methods). We find that the value of Moran's I of the active population distribution is mostly between 0.55 and 0.75 (Supplementary Table 3), implying that different cities have similar spatial autocorrelation patterns in terms of population distribution. Since the difference in the values of Moran's I between different cities is small, we cannot directly test the relationship between Moran's I and super-linear scaling through empirical data. Thus, we conduct a series of numerical simulations to generate point patterns with different Moran's Is (Supplementary Fig. 9). We generate $1 \times 10^5$ points under a two-dimensional Gaussian distribution with the mean $\mu = 0$ and the standard deviation $\sigma$ varying from 0.25 to 4. We then partition the space by $0.5 \times 0.5$ grid cells and calculate the interaction between each cell pair based on the gravity equation ($\gamma = 1$). For each $\sigma$, we run 30 simulations and take the average values of $\sigma$ and $\beta$. We highlight two simulation findings: 1) the simulated point distribution and gravity equation effectively resemble the super-linear scaling patterns and exponents (Fig.~\ref{fig:4}c). 2) $\beta$ increases monotonically as Moran's I increases, and the theoretical value $\beta = 7/6$ corresponds to Moran's I = 0.66 (Fig.~\ref{fig:4}d), the similar value to the empirical results of ten cities (Supplementary Table 3). All these findings point to a promising direction to study the in-depth connection between spatial patterns and scaling phenomena. 

\section*{Discussion}
In summary, we analyzed a diverse set of urban data, and find that cities exhibit robust intra-urban power-law scaling at the mesoscopic level: the infrastructure and socioeconomic activity satisfy sub- and super-linear exponents, respectively. Because the size of grid cells used here is somewhat arbitrary, we perform a sensitivity analysis by varying the cell size, and all conclusions are robust (Supplementary Table 2). Notably, the average intra-urban scaling exponents are consistent with previous cross-city results, providing direct empirical support to the hypothesis that cities are self-similar \cite{batty1994fractal} and manifest power-law scaling inside themselves as well. This finding also echoes the fractal nature of urban systems.

To explain the observed regularity and heterogeneity in the mesoscopic scaling phenomena, we provide a conceptual framework by decomposing spatial interactions into local and global effects. The sub-linear scaling of infrastructure volume can be derived through the local effect and is found to be jointly influenced by population and footprint areas. This sub-linearity also implies that urban spaces are `shared' by citizens, because if the area occupied per capita is constant, there will be no sub-linear phenomenon. Nevertheless, the shareability of different land types is different. In particular, we find that the per capita living area does not seem to change with city size (Supplementary Table 4). Detailed analysis of this phenomenon is beyond the scope of this study, but quantifying the scaling law between different land types and population size is worthy of further investigation.

The super-linear scaling is attributed to the city-wide interactions, which links urban scaling to human mobility. By adjusting the city-specific parameters $\alpha$, $\eta$, and $\gamma$, we give a better description of the real world, where the scaling exponents do not always appear symmetrically as $\beta = 1 \pm \delta$ for super- and sub-linear scaling predicted by previous models (Table~\ref{tab:2}). In particular, there is always a higher exponent for some super-linear scaling phenomena such as innovation and epidemic spreading \cite{bettencourt2007growth}; this may be primarily due to these phenomena being affected more by global interactions (a larger $\gamma$ or a more autocorrelated population distribution). 

From an application point of view, the revealed properties of intra-urban scaling are of considerable policy relevance. For instance, we can map granular socioeconomic activity or population distributions with only building data, which is available for many cities. Moreover, urban planners could assess the matching rate of facilities, socioeconomic activity, and population at a mesoscopic level within wities, thus guiding urban planning and renewal in the future.

It is important to note that, due to the accessibility of the dataset, we only present the results from ten large Chinese cities with high population density. Further research is needed to show whether the revealed patterns hold in other configurations, such as a spatially constrained city like Seattle or San Francisco, or a city whose growth has been largely uncontrolled, such as Los Angeles or Mexico City. Also, because our framework is minimal, it ignores various factors, such as transportation investment, policy, geographical barriers, all of which could affect the distribution of urban elements and the studied scaling phenomena. However, this paper provides an empirical and theoretical basis, where additional data and factors can be incorporated.

\section*{Methods}
\textbf{Population distribution dataset.}
The population distribution is estimated by a large-scale mobile phone dataset, which is provided by one location-based service provider in China. The mobile phone data have been used in our previous studies \cite{dong2017measuring, dong2019predicting}, and the population coverage of this dataset is shown in Supplementary Table 1. To protect user's privacy, we adopt a very rigorous protocol in this research. Firstly, all user IDs in our data are hashed and anonymized to ensure that one cannot associate the data to individual users. Secondly, all the researchers must follow a confidential agreement to use data for approved research. Thirdly, we use data aggregated into the grid cell for this study. To estimate the daytime and nighttime population distributions, we take the following steps: 

\begin{itemize}
    \item (i) Detecting stay point. For each anonymous individual, we have a series of geo-positiong points \{timestamp, longitude, latitude\}. A stay point is defined by a moving distance less than $d = 200$ m within a $t = 10$ min time threshold. As documented in our previous research \cite{dong2017measuring}, the stay points are robust when adjusting these thresholds within reasonable ranges. 
    \item (ii) Clustering. We cluster the stay points into different clusters using the DBSCAN algorithm \cite{ester1996density}. These clusters are defined as the stay locations. 
    \item (iii) Classification. We extract 28 features from the data (see Supplementary Table 5 for the main features). Then, we use Xgboost \cite{chen2016xgboost}, a supervised machine learning algorithm, to train two classifiers for the work and home location classification, respectively. The classification models are trained with a ground truth dataset, which contains $\sim 10^4$ users' self-reported home and work locations \cite{dong2019predicting}. The distributions of work and home locations are regarded as the daytime and nighttime population distributions, respectively. Figure~\ref{fig:1}bc present the spatial distributions of detected home and work locations of Beijing. 
\end{itemize}

To verify the accuracy of the results, we calculate the correlation between mobile phone data inferred home locations and the micro-census data of the year 2015 (the same year of our mobile phone dataset) at the district level. The $R^2$s of the linear regression ($\log MobilePhone = \beta \log Survey + \epsilon$) are 0.97 for Beijing and 0.98 for Shanghai, indicating that the mobile phone estimated population has good consistency with the survey data in terms of geographical distribution (Supplementary Fig. 10). The correlation between mobile phone data and official statistics has also been discussed in the studies of Estonia \cite{ahas2010using}, Portugal \cite{deville2014dynamic}, and France \cite{vanhoof2018assessing}.

\textbf{Building dataset.}
The building data were collected from one digital map in China. The geographical layouts of the buildings are presented in Supplementary Fig. 1 and Supplementary Fig. 11. We should note that since there is no ground truth for the building dataset, we cannot directly measure its quality. In Ref. \cite{US}, researchers from Microsoft track some metrics to measure the quality of a similar building dataset in the US. The IoU (intersection over union) of that test set is 0.85.

\textbf{Firm dataset.}
We collected firm registration record data from the registry database of the State Administration for Industry and Commercial Bureau of China. This dataset covers the registered information for firms in China, with attributes including firm name, year established, address, operation status, etc. We geocode firm addresses into longitude and latitude and then aggregate firms by grid cells of each city. Two limitations of the firm data should be noted: firstly, we only have registered address, which may not be the same as the operation address; and secondly, firm size (e.g., the number of employees or the revenue) is unreported in the raw dataset. 

\textbf{POI dataset.}
We collected POI data from \url{dianping.com}, the largest online rating website in China. The raw data include detailed locations of restaurants, shops, and service businesses (e.g., hair salon, photo studio), here we use points of restaurants and shops for our analysis. We note that the penetration rates of dianping.com in these two categories are high. For example, according to a report by Beijing Cuisine Association, there were 147,575 restaurants in operation at the end of 2016. In our dataset, we have 139,131, which covers 94.3\% of the total number of restaurants.

\textbf{Threshold.}
To make the results comparable across cities, we restrict all our data and analysis within the urban core area (the distance from the city center $\leq$ 15km for Beijing and $\leq$ 10km for the remaining cities. The coordinates of the city center are presented in Supplementary Table 1). To reduce the potential noise in the datasets, we further set four thresholds -- $10^{-2} \, \text{km}^2$ for footprint area, 1,000 for mobile phone estimated population, 2 for the number of firms and POIs -- to remove cells with values less than the thresholds. The number of cells used in the regression is shown in Supplementary Table 1.

\textbf{Grid cell.}
We transform the coordinate of each data point to a projected system (Gauss-Kruger) and build the grid system. For the grid cell division, we have two further explanations. The first is about the modifiable areal unit problem. With this grid style division, we can use different cell sizes to verify the robustness of the conclusions, which we have discussed in the Discussion section, Supplementary Table 2, and Supplementary Fig. 3. The second point is about a fundamental question -- how to define a city. Undoubtedly, a city is composed of a series of sub-units. According to the theory of fractal cities or hierarchical network-embedded cities, we have reason to find self-similar units within cities. This kind of grid cell division provides a basis for us to find such a unit. Specifically, the $2km \times 2km$ grid corresponds to the typical activity range of people's daily life, which is equivalent to a 15min living circle (people walk at a speed of 4-5 km/h). 

\textbf{Moran's I.}
To calculate Moran's I, we use the following formula:

\begin{equation}
    MI = \frac{n}{W} \frac{\sum_{i=1}^n \sum_{j=1}^n w_{ij} z_i z_j}{\sum_{i=1}^n z_i^2},
\end{equation}

\noindent where $n$ is the number of observations (grid cells in our case), $W$ is the sum of the weights $w_{ij}$ for all cell pairs in a city, $z_i = x_i - \bar{x}$ where $x$ is the active population size at location $i$ and $\bar{x}$ is the mean active population size in the city. Moran's I has a value from -1 to 1: -1 means perfect clustering of dissimilar values (i.e., perfect dispersion); 0 indicates no autocorrelation (i.e., perfect randomness); and 1 indicates perfect clustering of similar values (opposite of dispersion).

\subsection*{Acknowledgements}
We thank referees for providing constructive comments, Micheal Goodchild and seminar participants at Peking University for helpful discussions. This research was supported by the National Natural Science Foundation of China (nos. 41801299, 41625003, 41830645) and the China Postdoctoral Science Foundation (no. 2018M630026).

\subsection*{Author contributions}
L.D., Z.H., J.Z., and Y.L. designed research; L.D. performed research; L.D., J.Z., and Y.L. analyzed data; L.D. and Y.L. wrote the paper.

\subsection*{Data and code availability}
Data and code necessary to reproduce our results are available through \url{https://github.com/leiii/MesoScaling}.

\subsection*{Competing interests}
The authors declare no competing interests and no financial conflict of interest as well.


\end{document}